\documentclass{webofc}

\usepackage[varg]{txfonts}   
\usepackage{xpatch}
\xpretocmd{\cref}{Eq.~}{}{}
\usepackage{cleveref}
\begin{document}
\title{Sterile Neutrino. }
\subtitle{A short introduction.}

\author{\firstname{Dmitry V.} \lastname{Naumov\inst{1}\fnsep\thanks{\email{dnaumov@jinr.ru}}}}
\institute{Joint Institute for Nuclear Research}
\abstract{%
This is a pedagogical introduction to the main concepts of the sterile neutrino - a hypothetical particle, coined to resolve some anomalies in neutrino data and retain consistency with observed widths of the $W$ and $Z$ bosons.
We briefly review existing anomalies and the oscillation parameters that best describe these data.

We discuss in more detail how sterile neutrinos can be observed, as well as the consequences of its possible existence.
In particular, we pay attention to a possible loss of coherence in a model of neutrino oscillations with sterile  neutrinos, where this effect might be of a major importance with respect to the 3$\nu$ model. 

The current status of searches for a sterile neutrino state is also briefly reviewed. 
}
\maketitle
\section{Introduction}
\label{sec:intro}
There are three generations of leptons in the Standard Model (SM)
\begin{equation}
\label{eq:leptons}
\left(
\begin{array}{c}
\nu_1\\
e  \\
\end{array}
\right)_L, 
\hspace{0.1\linewidth}
\left(
\begin{array}{c}
\nu_2\\
\mu  \\
\end{array}
\right)_L, 
\hspace{0.1\linewidth}
\left(
\begin{array}{c}
\nu_3\\
\tau  \\
\end{array}
\right)_L 
\end{equation}
grouped in SU(2)$_L$ doublets. 
The sub-index $L$ indicates that the quantum fields $\nu_i$ ($i=1,2,3$) and $\ell_\alpha$ ($\alpha=e,\mu,\tau$) are eigenstates of the $P_L=\frac{1}{2}(1-\gamma_5)$ left-handed helicity operator.

The fields $\nu_i$ and $\ell_\alpha$ have definite masses and they obey the Dirac equation.
There are special linear combinations of $\nu_i$ fields known as flavor neutrinos $\nu_\alpha$
\begin{equation}
\label{eq:flavor_vs_mass}
\left(
\begin{array}{c}
\nu_e\\
\nu_\mu\\
\nu_\tau\\
\end{array}
\right)
=
V
\left(
\begin{array}{c}
\nu_1\\
\nu_2\\
\nu_3\\
\end{array}
\right),
\end{equation}
where $V$ is $3\times 3$ unitary  Pontecorvo-Maki-Nakagawa-Sakata (PMNS) matrix~\cite{Pontecorvo:1957qd,Maki:1962lba}.

The  fields of flavor neutrino $\nu_\alpha$ have definite lepton numbers $L_\alpha$, ($\alpha=e,\mu,\tau$).
If the masses of $\nu_i$ are all different, the  flavor neutrino field $\nu_\alpha$ does not obey the Dirac equation, nor is the lepton number is conserved.
Therefore, there is not much sense in using the flavor neutrino fields, as they are not fundamental objects of the SM.
Respectively, they can be abandoned in any consideration.

In the SM the interactions of leptons with the $W$-boson mix all generations of the former, as can be seen from the corresponding  Lagrangian term
\begin{equation}
\label{eq:mixing_W}
\mathcal{L} = -\frac{g}{\sqrt{2}} \sum_{\alpha=e,\mu,\tau}\sum_{i=1,2,3} V_{\alpha i} \overline{\nu}_i\gamma^\mu P_L\ell_\alpha W_\mu + \text{h.c.},
\end{equation}
where $V_{\alpha i}$ is the matrix element of the PMNS matrix $V$, $\overline{\nu}_i$, $\ell_\alpha$, $W_\mu$ are quantum fields of the neutrino with mass $m_i$, lepton of flavor $\alpha$ and $W$-boson, respectively and  $\gamma^\mu$ is a Dirac $4\times 4$ matrix.

The smallness of the masses of neutrinos and their mixture in interactions with $W$-boson and charged leptons from different generations give rise to a spectacular quantum effect observed at macroscopic scales -- {\em oscillation of lepton flavor}, or -- {\em neutrino oscillation}~\cite{Fukuda:1998mi,Ahn:2002up,Ahmad:2001an,Ahmad:2002jz,Abe:2008aa,An:2012eh,An:2016ses}.
This effect manifests itself as a quasi-periodic probability to observe charged leptons $\ell_\alpha$ and $\ell_\beta$ in, respectively, the source and the detector of neutrino.

Simplifying the consideration to only two neutrino types $\nu_1$ and $\nu_2$, the corresponding probability in the plane wave model reads
\begin{equation}
\label{eq:osc_prob_planewave}
\begin{aligned}
P_{\alpha\beta} & = \sin^22\theta\sin^2\frac{\Delta m^2 L}{4p}, \alpha\ne \beta,\\
P_{\alpha\alpha}&=1-\sin^22\theta\sin^2\frac{\Delta m^2 L}{4p}.
\end{aligned}
\end{equation}  
In~\cref{eq:osc_prob_planewave} $\theta$ is the mixing angle, parameterizing the PMNS matrix $V$ in~\cref{eq:flavor_vs_mass}, $\Delta m^2 = m_2^2-m_1^2$, $L$ is the distance between the source and the detector and $p$ is the absolute value of neutrino three-momentum.
In a two-neutrino model the oscillation probability is strictly periodic as a function of $L/p$.
The corresponding oscillation length 
\begin{equation}
L^\text{osc}=4\pi p/\Delta m^2 = 2.48 \text{km} \frac{p}{\text{MeV}}\frac{10^{-3}\text{eV}^2}{\Delta m^2}
\end{equation} is of macroscopic magnitude for observed $\Delta m^2$ and typical neutrino momenta.

The plane-wave model of neutrino oscillation being used elsewhere is not self-consistent and leads to a number of paradoxes~\cite{Akhmedov:2009rb}.

A consistent model adopts wave-packets.
The oscillation probability in~\cref{eq:osc_prob_planewave} is modified (see~\cite{Naumov:2010um,An:2016pvi} and references therein) and in the wave-packet model it depends on three more parameters having dimension of length -- coherence($L^\text{coh}$),  dispersion ($L^\text{d}$) lengths and spatial size $\sigma_x$ of the neutrino wave-packet (reciprocal to its  three-momentum dispersion $\sigma_p$).
The interference term in~\cref{eq:osc_prob_planewave} is suppressed by 
\begin{equation}
\label{eq:incoherence_factor}
e^{-\left(1+\left(L/L^d\right)^2\right)^{-1}\left(L/L^\text{coh}\right)^2}e^{ -\left(\sqrt{2}\pi\sigma_x/L^\text{osc}\right)^2}
\end{equation} 
factors, where
\begin{equation}
\label{eq:coh_disp_leghts}
L^\text{coh} = \frac{2\sqrt{2}p^2}{\sigma_\text{p}\Delta m^2},\quad 
L^\text{d}    = \frac{p^3}{\sigma_{\text{p}}^2 \Delta m^2},\quad 
\sigma_x      = \frac{1}{2\sigma_p}.
\end{equation}
One consequence of the wave-packet model of neutrino oscillation is that for $L\gg L^\text{coh}$ the neutrino is  not in a coherent superposition and the probability of observation $P_{\alpha\beta} $ depends neither on distance, nor on momentum.  
For any realistic assumption about $\sigma_p$ neutrinos traveling astrophysical distances are incoherent.

Real analyses of neutrino oscillation data use a three-neutrino model.
Analyses of solar, atmospheric, reactor and accelerator neutrino data yield 
\begin{equation}
\label{eq:observed_osc_pars}
\Delta m^2_{12} \simeq 7.5\cdot 10^{-5}\text{eV}^2, \quad |\Delta m^2_{32}|\simeq 2.5\cdot 10^{-3}\text{eV}^2.
\end{equation}

\section{Anomalies in neutrino data}
\label{sec:anomalies}
The motivation for proposing a sterile neutrino state is driven by the existence of some anomalies in neutrino data which cannot be described by a three-neutrino model with values of $\Delta m^2_{12}$ and $|\Delta m^2_{32}|$ given by~\cref{eq:observed_osc_pars}.

There are two groups of the corresponding anomalies seen as {\em appearance} and {\em disappearance} of neutrinos.
\subsection{Appearance and disappearance data}
Appearance data include (i) LSND and (ii) MiniBooNE observations. 

LSND observed an excess of $\overline{\nu}_e$ in their study of decays $\mu^+\to e^++\nu_e+\overline{\nu}_\mu$ of positively charged muons at rest~\cite{Aguilar:2001ty}. 
The mean neutrino energy was about 30 MeV. 
The statistical significance of the excess was about three standard deviations.  

MiniBooNE, with a $\nu_\mu (\overline{\nu}_\mu)$ beam having neutrino energy of $600 (400)$ MeV and the same $L/p$ as in LSND, observed an excess of $\nu_e$ and $\overline{\nu}_e$ with a significance of about $4.8$ standard deviations~\cite{Aguilar-Arevalo:2018gpe}.

Disappearance data include (i) an about 13\% deficit of $\nu_e$ and $\overline{\nu}_e$ from calibration sources of SAGE~\cite{Abdurashitov:1998ne,Abdurashitov:2005tb} and GALLEX~\cite{Bahcall:1994bq} experiments and (ii) a 6\% deficit of $\overline{\nu}_e$  from reactors when compared to a calculation~\cite{Huber:2011wv,Mueller:2011nm}.

We note that none of these anomalies show a distinct $L/p$ oscillation dependence predicted by~\cref{eq:osc_prob_planewave}.  
\subsection{Interpretation of anomalies within the hypothesis of neutrino oscillation}
\label{sec:interpretation}
If these anomalies are interpreted as due to neutrino oscillation then $\Delta m^2\simeq (1-2)$ eV${}^2$ and $\sin^22\theta\simeq (0.1-0.2)$ are best to describe these observations.
The required $\Delta m^2\simeq (1-2)$ eV${}^2$ does not fit into the 3$\nu$ framework with measured values of $\Delta m^2$ given by~\cref{eq:observed_osc_pars}.

A straightforward extension of the lepton sector of the SM by adding an additional doublet of leptons is impossible because of the measured widths of $W$ 
\begin{equation}
\label{eq:W_width}
	\Gamma(W\to all) = \left(2085\pm 2.1\right)\text{\; MeV}
\end{equation}	   
and of $Z$ bosons
\begin{equation}
\label{eq:Z_width}
\Gamma(Z\to all) = \left(2495\pm 2.3\right)\text{\; MeV}.
\end{equation}

Each new neutrino (with neutrino mass less than $m_Z/2$) from the SU(2)$_L$ doublet adds to the widths of $W$ and $Z$ the following contributions $\Gamma(W\to \ell \nu)\simeq 226 $ MeV and $\Gamma(Z\to \overline{\nu} \nu)\simeq 166$ MeV which are much larger than the experimental uncertainties in~\cref{eq:W_width,eq:Z_width}.

These are essential, minimally required ingredients to review the concept of sterile neutrino.

\section{Concept of Sterile Neutrino}
\subsection{Masses of fermions in the SM }
The main trick to create a sterile neutrino state is to add to the SM a neutrino field without adding the fourth left-handed fields of leptons.
In order to explain it, let us recall how the masses of fermions appear in the SM.
We simplify our consideration by examining only the Dirac-type particles in order to keep the analysis simple and clear.

A fundamental idea of the SM is a requirement of gauge invariance of the Lagrangian under a particular symmetry, SU(2)${}_L\times$U(1).
All fermions  in the SM are massless because the corresponding mass-term is not gauge invariant.

To illustrate this statement let us consider first the mass-term of only one generation of charged leptons
\begin{equation}
\label{eq:mass_term}
-m \left(\overline{\ell}_L \ell_R + \text{h.c.}\right)
\end{equation}
\Cref{eq:mass_term} is not gauge invariant because the left-handed $\ell_{L}$ and right-handed  $\ell_{R}$ fields have different gauge symmetries: SU(2)${}_L\times$U(1) and U(1), respectively.

The masses of fermions are acquired via the so-called Yukawa interactions of a scalar field with two fields of fermions.
Originally, this interaction was proposed to explain an attractive potential of two nucleons interacting with each other by exchange of a scalar massive field -- the pion.
In the SM the same terminology is used to describe interactions of fermions with the only fundamental scalar field of the model -- the Higgs field $\varphi$.
The Yukawa term in the Lagrangian for charged leptons reads 
\begin{equation}
\mathcal{L}_Y = - \lambda \left(\overline{\nu}_L, \overline{\ell}_L\right)
\left(
\begin{array}{c}
0\\
\varphi
\end{array}
\right)\ell_R + \text{h.c.}
\end{equation}
The replacement $\varphi\to v + \frac{\chi}{\sqrt{2}}$, where $v$ is a non-zero vacuum expectation value (VEV) of the Higgs field, generates  the mass term in~\cref{eq:mass_term} with $m=\lambda v$ 
and a term describing an interaction of fermions with an excitation $\chi$ of the Higgs field above VEV. 

The masses of the neutrinos are generated in a similar way replacing 
$\left(
\begin{array}{c}
0\\
\varphi
\end{array}
\right)$ by 
$
\left(
\begin{array}{c}
	-\varphi^c\\
	0
\end{array}
\right)
$, where $\varphi^c$ is a charge-conjugated Higgs field, and replacing $\ell_R$ by $\nu_R$.
The right-handed neutrino field $\nu_R$ does not interact with $W$ and $Z$, and thus is often called  ''sterile''. 
This is not the sterile field one needs in interpretations of anomalies in neutrino data mentioned in Sec.~\ref{sec:anomalies}.

In order to generalize our consideration of the case of generations of leptons one should add three right-handed fields for charged leptons $\ell_{eR}$, $\ell_{\mu R}$, $\ell_{\tau R}$ and three  right-handed fields for neutrinos $\nu_{eR}$, $\nu_{\mu R}$, $\nu_{\tau R}$.
The most general Yukawa Lagrangian gives non-diagonal  terms for charged leptons and neutrinos
\begin{equation}
\label{eq:lagrangian_mass_terms}
-m^\ell_{\alpha\beta}\left(\overline{\ell}_{\alpha L}\ell_{\beta R}+\text{h.c.}\right)-m^\nu_{\alpha\beta}\left(\overline{\nu}_{\alpha L}\nu_{\beta R}+\text{h.c.}\right).
\end{equation}
$m^\ell$ and $m^\nu$ are  non-diagonal matrices with matrix elements $m^\ell_{\alpha\beta}$ and $m^\nu_{\alpha\beta}$, respectively.

Both terms in~\cref{eq:lagrangian_mass_terms} must be diagonalized in order to be interpreted as mass-terms.
The diagonalization can be done with the help of four matrices $U^\ell_L$, $U^\ell_R$, $U^\nu_L$ and $U^\nu_R$, rotating $\ell_L$, $\ell_R$, $\nu_L$ and $\nu_R$, respectively, where $\ell^T=\left(\ell_e,\ell_\mu,\ell_\tau\right)$ and $\nu^T=\left(\nu_e,\nu_\mu,\nu_\tau\right)$.

The rotation matrices  make $U^{\ell\dagger}_L m^\ell U^\ell_R$ and $U^{\nu\dagger}_L m^\nu U^\nu_R$ diagonal.
As a result, the fields with definite masses and belonging to different SU(2)${}_L$ doublets   mix in their interactions with $W$ boson as shown in~\cref{eq:mixing_W}. 
The corresponding mixing PMNS-matrix reads $V=U^{\ell\dagger}_L U^\nu_L$.
This matrix  should not be attributed, as done often in the literature, to neutrino fields.
Instead, $V$ is a mixing matrix of both charged leptons and neutrinos.

The reviewed mass-generation mechanism used fields for three charged leptons and three neutrinos. 
\subsection{Four right-handed neutrinos and  three left-handed doublets}
The sterile neutrino emerges if there are four  right-handed neutrino and still three left-handed doublets.
Let us label the fourth right-handed neutrino field as $\nu^s_R$.
Assuming that this new field interacts with the Higgs and left-handed neutrino fields in Yukawa interactions, one arrives at non-diagonal terms like in~\cref{eq:lagrangian_mass_terms} in which $m^\nu$ is replaced by a $4\times 4$ matrix and an additional fourth left-handed field is  constructed as $\left(\nu^s_R\right)^c$.
Making the diagonalization, one finds that $3\times 3$ mixing matrix $V$ is replaced by a $4\times 4$ matrix
\begin{equation}
\label{eq:unitary_matrix_4x4}
\left(
\begin{array}{cc}
V_{3\times 3} & K_{3\times 1}\\
U_{1\times 3} & M_{1\times 1},
\end{array}
\right)
\end{equation}
in which the dimension of each sub-block is displayed.
The unitarity of this matrix yields the following relationships
\begin{equation}
\label{eq:unitarity}
\begin{aligned}
V^\dagger V+U^\dagger U & =1_{3\times 3}, \quad K^\dagger K+M^\dagger M & =1_{1\times 1},  \\
V V^\dagger+K K^\dagger & =1_{3\times 3},\quad  U U^\dagger+M M^\dagger & =1_{1\times 1},\\
V^\dagger K+U^\dagger M & =0_{3\times 1},  \quad V U^\dagger +KM^\dagger  & =0_{3\times 1}.
\end{aligned}
\end{equation}
Grouping flavor neutrino fields into $\nu^f=\left(\nu_e,\nu_\mu,\nu_\tau\right)^T$ and massive neutrino fields into $\nu^m=\left(\nu_1,\nu_2,\nu_3\right)^T$, the relationship in~\cref{eq:flavor_vs_mass} is generalized as
\begin{equation}
\label{eq:flavor_vs_mass_4nu}
\left(
\begin{array}{c}
	\nu^f_L\\
	(\nu^s_R)^c
\end{array}
\right)
=
\left(
\begin{array}{cc}
	V_{3\times 3} & K_{3\times 1}\\
	U_{1\times 3} & M_{1\times 1}
\end{array}
\right)
\left(
\begin{array}{c}
	\nu^m_L\\
	\nu^4_L
\end{array}
\right).
\end{equation}
The left-handed neutrino field $\nu^s_L\equiv (\nu^s_R)^c$ being made of a sterile right-handed field, remains sterile in terms of interactions with $W$ and $Z$-bosons.
At the same time fields $\nu^m_L$ and $\nu^4_L$ do interact with $W$ and $Z$ bosons.

What can be said about widths of $W$ and $Z$ bosons in the presence of new fields $\nu^4_L$ and $\nu^s_L$?
The corresponding amplitude, assuming coherence of the sterile neutrino state and using~\cref{eq:unitarity}, reads for $W$
\begin{equation}
\label{eq:amplitude_sterile_W}
\mathcal{A}^W(\nu_s + W^-\to \ell^-_\alpha)   = \sum_{i=1}^3 U^*_{1i}V_{\alpha i}\mathcal{A}^W_i + M^*_{11}K_{\alpha 1}\mathcal{A}^W_{4} 
\simeq (VU^\dagger + KM^\dagger)_{\alpha 1}\mathcal{A}^W_0 = 0,
\end{equation}
where $\mathcal{A}^W_i, \mathcal{A}^W_4$ are interaction amplitudes for massive neutrinos and $\mathcal{A}^W_0$ is the corresponding amplitude assuming zero neutrino mass. 

Similarly, one can show that for the $Z$ boson, within the same assumptions, the amplitudes read
\begin{equation}
\label{eq:amplitude_sterile_Z}
\begin{aligned}
&\mathcal{A}^Z(\nu_s + Z^0\to \nu_i) = \sum_j U^*_{1j}\mathcal{A}^Z_{ji} + M^*_{11}\mathcal{A}^Z_{4i}
\simeq \mathcal{A}_0^Z\left(V^\dagger\left[VU^\dagger+KM^\dagger\right]\right)_{i1}=0,\\
&\mathcal{A}^Z(\nu_s + Z^0\to \nu_4) = \sum_j U^*_{1j}\mathcal{A}^Z_{j4} + M^*_{11}\mathcal{A}^Z_{44}
\simeq \mathcal{A}_0^Z\left(K^\dagger\left[VU^\dagger+KM^\dagger\right]\right)_{11}=0.
\end{aligned}
\end{equation} 
\Cref{eq:amplitude_sterile_W,eq:amplitude_sterile_Z} show that the interaction amplitudes of a sterile neutrino state with $W$ and $Z$ are both vanishing.
Therefore, decays of  $W$ and $Z$ involving sterile neutrinos should also vanish in the end.
We made an important assumption that a coherent superposition of neutrino states $|\nu_i\rangle$ and $|\nu_4\rangle$ making up the sterile state $|\nu_s\rangle=\sum_{i=1}^3U^*_{1i}|\nu_i\rangle+M^*_{11}|\nu_4\rangle$ can be produced.

As mentioned in~\cref{sec:intro} use of states with definite momentum is an approximation which fails in describing neutrino oscillation phenomenon. 
A more consistent approach is based on using the wave-packet model which necessarily imposes a non-zero incoherency in the production of states with different masses. 
The term suppressing the coherence of neutrino states is the second term in~\cref{eq:incoherence_factor}.
The coherence is suppressed if the wave-packet spatial dimension $\sigma_x$ is comparable to or larger than the oscillation length $L_\text{osc}$.
Let us note that in the plane wave model $\sigma_x=\infty$.

What happens to $W$ and $Z$ widths if  the $|\nu_s\rangle$ state cannot be produced as a coherent superposition of $|\nu_i\rangle$ and $|\nu_4\rangle$ states?
Let us focus only on the $Z$-boson decay width.
One can observe that the diagonal in the SM vertex 
\begin{equation}
\label{eq:SM_Z_nunu}
-\frac{g}{\cos\theta_W}\sum_{\alpha=e,\mu,\tau}\overline{\nu}_{\alpha, L}\gamma^\mu\nu_{\alpha,L} Z_\mu=-\frac{g}{\cos\theta_W}\sum_{i=1}^3\overline{\nu}_{i, L}\gamma^\mu\nu_{i,L} Z_\mu
\end{equation}
of $Z$-boson interaction  with neutrino  is no longer diagonal in an extension of the SM with sterile neutrinos.
Keeping only essential factors, this vertex now reads
\begin{equation}
\label{eq:Z_vertex_with_sterile}
\sum_\alpha\overline{\nu}_\alpha\nu_\alpha \to 
 \sum_{i,j}\left(V^\dagger V\right)_{ij} \overline{\nu}_i\nu_j+ \sum_i\left(V^\dagger K\right)_{i1}\overline{\nu}_i\nu_4 
+ \sum_i\left(K^\dagger V\right)_{1i}\overline{\nu}_4\nu_i+\left(K^\dagger K\right)_{11}\overline{\nu}_4\nu_4. 
\end{equation}
The first term in~\cref{eq:Z_vertex_with_sterile} suggests that the decay width $Z\to \overline{\nu}_i \nu_j$ is proportional to   $|(V^\dagger V)_{ij}|^2$ and therefore, summation over $i,j$ yields
\begin{equation}
\sum_{i,j} \Gamma(Z\to \overline{\nu}_i \nu_j) \propto \text{Tr}\left(V^\dagger V V^\dagger V\right).
\end{equation} 
A similar calculation can be applied to all other possible final states.
Therefore, the width $\sum_{i,j} \Gamma(Z\to \overline{\nu}_i \nu_j)  + \sum_{i} \Gamma(Z\to \overline{\nu}_i \nu_4)  + \sum_{i} \Gamma(Z\to \overline{\nu}_4 \nu_i)+ \Gamma(Z\to \overline{\nu}_4 \nu_4)$ of a $Z$-boson decaying into neutrino and anti-neutrino
is proportional to
\begin{equation}
\label{eq:Z_decay_width}
\begin{aligned}
& \text{Tr}\left(V^\dagger V V^\dagger V+V^\dagger K K^\dagger V + K^\dagger V V^\dagger K+K^\dagger K K^\dagger K \right)\\
&= \text{Tr}\left(V V^\dagger V V^\dagger +VV^\dagger K K^\dagger  + K^\dagger V V^\dagger K+K K^\dagger K K^\dagger  \right)\\
&= \text{Tr}\left(V V^\dagger + K K^\dagger\right)^2=3,
\end{aligned}
\end{equation}
where for the last equality we used first unitarity relation in the second line of~\cref{eq:unitarity}.
Thus, according to~\cref{eq:Z_decay_width}, the  expected width of a $Z$ boson decaying into all possible combinations of four neutrino and anti-neutrino states is determined by three generations of leptons avoiding inconsistency with the observed decay width of $Z$ boson.

In order to get a simpler understanding of how this magic happened, it is instructive to consider only one lepton generation, rather than three, for example $\left(\nu_{e,L},e_L\right)$.
Let us now add two right-handed neutrino fields and a Yukawa interaction term which should be diagonalized, similar to considerations with three generations of leptons.
Instead of~\cref{eq:flavor_vs_mass_4nu} one gets
\begin{equation}
\label{eq:flavor_vs_mass_2nu}
\left(
\begin{array}{c}
\nu_e\\
\nu_s
\end{array}
\right)
=
\left(
\begin{array}{cc}
\cos\theta & \sin\theta\\
-\sin\theta & \cos\theta
\end{array}
\right)
\left(
\begin{array}{c}
\nu_1\\
\nu_2
\end{array}
\right).
\end{equation}
The flavor part ($\overline{\nu}_e\nu_e$) of the Lagrangian given by~\cref{eq:SM_Z_nunu} for $Z$ boson interaction with $\nu_e$ and $\overline{\nu}_e$ is replaced by
\begin{equation}
\label{eq:Z_2nu}
\cos^2\theta\overline{\nu}_1\nu_1+\sin\theta\cos\theta\left(\overline{\nu}_2\nu_1+\overline{\nu}_1\nu_2\right)+\sin^2\theta^2\overline{\nu}_2\nu_2.
\end{equation}
Therefore, the width of $Z$ boson decaying into any of $\overline{\nu}_i\nu_j, (i,j=1,2)$ is proportional to
\begin{equation}
\label{eq:Z_2nu_width}
\underbrace{\cos^4\theta}_{\overline{\nu}_1\nu_1}+\underbrace{\sin^2\theta\cos^2\theta}_{\overline{\nu}_2\nu_1}+\underbrace{\sin^2\theta\cos^2\theta}_{\overline{\nu}_1\nu_2}+\underbrace{\sin^4\theta}_{\overline{\nu}_2\nu_2}
 =\left(\cos^2\theta+\sin^2\theta \right)^2=1.
\end{equation}
Thus, we find that the corresponding  decay width is proportional to one, while there are two massive neutrinos $\nu_1$ and $\nu_2$ both interacting with $Z$.
The original coupling of $Z\overline{\nu}_e\nu_e$ vertex is redistributed over four combinations of two fields conserving the total ''strength'' of the interaction as illustrated by~\cref{eq:Z_2nu,eq:Z_2nu_width}.

Let us briefly summarize key elements of the concept of sterile neutrinos.

(i)  Assume a  disparity between the numbers of  neutrino fields interacting (active) and non-interacting (inert) directly with $W$ and $Z$ bosons in the SM.  For example, assuming Dirac neutrinos, there are three left-handed interacting and four (or more) right-handed  non-interacting with $W$ and $Z$, neutrino fields.

(ii) Assume a mechanism of mixing 	active and inert neutrino fields in a generally non-diagonal mass-term. An example of such mechanism is the Yukawa interaction. Other mechanisms also exist.

(iii) The sterile neutrino field emerges after the diagonalization of the mass-term. The sterile field is a superposition of at least four massive neutrino fields with nearly zero interaction amplitude with  $W$ and $Z$. This amplitude corresponds to a coherent superposition of all four neutrino states.

(iv) The widths of $W$ and $Z$ bosons decaying into any possible combination of neutrino and anti-neutrino states is proportional to the number of active neutrino fields (three in the SM).
\subsection{How sterile neutrino state can be observed}
(i) In neutrino oscillation as a deficit of the event rate and as $L/p$ pattern in both appearance and disappearance channels. 
Remarkably, these effects are expected for both charged  and neutral currents.
This is in contrast to neutrino oscillation without the sterile neutrino, in which only charged currents display an oscillatory pattern and the event rates in neutral currents remains unchanged.

(ii) Since the $4\times 4$ matrix in~\cref{eq:unitary_matrix_4x4} is unitary, its $3\times 3$ sub-block $V$, which is the PMNS matrix, must be non-unitary unless $K$ and $U$ non-diagonal sub-blocks are identically zero.
In the latter case, the oscillation to a sterile neutrino state is impossible and no event rate deficit due to sterile neutrinos can be expected.
Therefore, a measurement of unitarity of $V_{3\times 3}$ is a direct probe of the existence of sterile neutrinos.

(iii) Beta-decays of tritium provide measurements of  
\begin{equation}
m_\beta=\left(\sum_{i=1} |V_{ei}|^2 m_i^2\right)^{1/2},
\end{equation}
sensitive to $m_4$ and to $|V_{e4}|^2$.

(iv) Neutrino-less beta decays of unstable nuclei, if the neutrino is a Majorana particle, provide  measurements of
\begin{equation}
m_{\beta\beta} = \left|\sum_i V^2_{ei} m_i\right|,
\end{equation}
sensitive to $m_4$ and to $V_{e4}$.

(v) In cosmology, $\nu_4$ is a relativistic degree-of-freedom in primordial plasma, which is an observable. 
The sterile neutrino state affects Big-Bang-Nucleosynthesis because a larger number of neutrino species means a faster expansion rate of the Universe.
There are many other ways in which the sterile neutrino impacts the evolution of the Universe. 
For example, an additional $\nu_4$ field would add to a contribution of relic neutrinos into the energy density of the Universe.
This contribution is determined by $\sum_i m_i$. 
\subsection{Loss of coherence for sterile neutrino}
The argument of the second exponential in~\cref{eq:incoherence_factor} is responsible for the coherency of neutrino states at production and detection.
It is not at all guaranteed that any combination of $|\nu_i\rangle$ could be coherently produced or detected, while this is usually silently assumed in a plane wave model.
These states would be coherent if $\left(\sqrt{2}\pi\sigma_x/L^\text{osc}\right)^2\ll 1$.
This condition is relativistically invariant, which can be seen writing 
\begin{equation}
\label{eq:coherence_suppression_factor}
\left(\sqrt{2}\pi\sigma_x/L^\text{osc}\right)^2 = \frac{1}{4}\left(\frac{\Delta m^2}{\sigma_{m^2}}\right)^2,
\end{equation}
where $\sigma_{m^2}=2\sqrt{2}p\sigma_p$ can be interpreted as uncertainty of measurement of $m^2$.
Thus, the coherency of $|\nu_i\rangle$ states is possible if the uncertainty  in the determination of $m^2$ is much larger than $\Delta m^2$.
This is in agreement with the principles of quantum physics -- if an intermediate state cannot be determined, one should sum amplitudes with all possible intermediate states, yielding interference terms in the absolute value squared of the total amplitude.

Consider neutrinos from pion decays. This reaction is important for atmospheric and accelerator neutrinos.
Since $\sigma_{m^2}$ is a relativistic invariant one can make an estimate of this quantity in the pion's rest-frame, taking $p=29.8$ MeV and $\sigma_p=\Gamma_\pi=2.5\cdot 10^{-8}$ eV, where $\Gamma_\pi$ is the decay width of $\pi$ meson in its rest-frame.
Therefore, one can estimate $\sigma_{m^2} = 2.1$ eV${}^2$.
This value is much larger than the observed $\Delta m^2$ given by~\cref{eq:observed_osc_pars} for a three neutrino model.
Taking the largest  $|\Delta m^2_{32}|\simeq 2.5\cdot 10^{-3}\text{eV}^2$, one can see that $1/4 \left(2.5\cdot 10^{-3}/2.1\right)^2\simeq 3.5\cdot 10^{-7}$ and the second exponential in~\cref{eq:incoherence_factor} is very close to unity, which means no significant suppression of the interference term.
A posteriori, this calculation confirms the assumption of neutrino coherence made in the plane wave model.

Taking $\Delta m^2\simeq (1-2)$ eV${}^2$ required to explain anomalies in neutrino data briefly reviewed in~\cref{sec:anomalies}, one can observe that the factor in~\cref{eq:coherence_suppression_factor} is approximately equal to $ (0.06-0.22)$ and the second exponential in~\cref{eq:incoherence_factor}  takes values $0.94-0.8$, sizably affecting the oscillation pattern.
For example, for $\Delta m^2\simeq 3.5$ eV${}^2$ the oscillation amplitude is suppressed by 50\%.
Correspondingly, an erroneous statement about the mixing angle of sterile neutrinos could be drawn in a plane wave model in which these suppressions are ignored.
A careful analysis of neutrino oscillation data requires a wave packet model.

As a side remark, let us note that  arguments similar to these considerations explain why charged leptons do not oscillate, their $\Delta m^2$  is much lager than $\sigma_{m^2}$.
Thus, the interference terms vanish~\cite{Akhmedov:2007fk}.
\subsection{Confusions in terminology}
One might be confused by the use of the same terminology ''sterile neutrino'' with different meaning by physicists working with neutrino oscillations and by cosmologists.

For the former ''sterile neutrino'' means a coherent superposition of mass eigenstates $|\nu_i\rangle$ like in our notation $|\nu_s\rangle=\sum_{i=1}^3U^*_{1i}|\nu_i\rangle+M^*_{11}|\nu_4\rangle=\sum_{i=1}^4V^*_{si}|\nu_i\rangle$.
Cosmologists often use ''sterile neutrino'' to refer to the fourth state $|\nu_4\rangle$ silently assuming that $|V_{\alpha 4}|^2\ll 1$.
 
\subsection{Current status and perspectives}
A world-wide research program is carried out examining the possible existence of a sterile neutrino state.
Here we very briefly review the current status suggesting an interested reader to follow dedicated reviews~\cite{Dentler:2018sju,Gariazzo:2017fdh}.

In 2018 there are several hints in favor of the existence of sterile neutrinos, and there is a bulk of data excluding possible parameter space for this still hypothetical particle.

A deficit of reactor $\overline{\nu}_e$ with respect to calculations~\cite{Huber:2011wv,Mueller:2011nm} is known now as the ''reactor antineutrino anomaly''.
Its interpretation as due to $\overline{\nu}_e\to\overline{\nu}_s$ oscillation with $\Delta m^2=2.3$ eV${}^2$ and $|V_{e4}|^2=0.14$ was addressed by a number of experiments.
DANSS~\cite{Alekseev:2018efk} and NEOS~\cite{Ko:2016owz} excluded the best-fit parameters of oscillation model with sterile neutrino, but their data favors at about $3\sigma$ confidence level $\Delta m^2=1.3$ eV${}^2$ and $|V_{e4}|^2\approx 0.1$~\cite{Dentler:2018sju}.
Daya Bay measured rates of  reactor $\overline{\nu}_e$ due to two dominant isotopes ${}^{235}\text{U}$ and ${}^{239}\text{Pu}$~\cite{An:2017osx,Adey:2018qct}.
It was found that the observed deficit is caused mainly by a larger model contribution of ${}^{235}\text{U}$, while the expected and measured rates of $\overline{\nu}_e$  due to ${}^{239}\text{Pu}$ are found to be consistent.
A model with sterile neutrinos suggesting equal deficit for any nuclear isotope is excluded by Daya Bay data at 2.6$\sigma$.

Appearance data from LSND~\cite{Aguilar:2001ty} and MiniBooNE~\cite{Aguilar-Arevalo:2018gpe}, interpreted as due to sterile neutrino oscillation, is in strong tension with recent disappearance data from MINOS/MINOS+~\cite{Adamson:2017uda}, NOvA~\cite{Adamson:2017zcg}, IceCube~\cite{TheIceCube:2016oqi}.
The compatibility of appearance and disappearance datasets is less than $2.6\cdot 10^{-6}$~\cite{Dentler:2018sju}. 
Therefore, the sterile neutrino interpretation of LSND and MiniBooNE anomalies is  unlikely.

Cosmology provides other strong constraints on the existence of sterile neutrino.
After a large number of collisions in early plasma the coherence of massive neutrinos would be lost and all four neutrino species $\nu_i, (i=1\dots 4)$ will be in a thermal equilibrium if $|V_{\alpha 4}|^2$ is not vanishingly small.
An additional relativistic degree of freedom is disfavored by  constraints from Big Bang Nucleosynthesis~\cite{Cyburt:2015mya} and from recombination epoch~\cite{Akrami:2018vks}.
The sum of neutrino masses $\sum_{i}m_i$ is significantly constrained by the Cosmic Microwave Background and structure formation data, which disfavor extra neutrino species with masses larger than 0.3 eV.

\section{Summary}
We reviewed main concepts of sterile neutrinos -- a yet hypothetical particle, coined to resolve some anomalies in neutrino data.
In the framework of an extension of the SM, a sterile neutrino field is a superposition of fields of massive neutrinos.
The corresponding interaction amplitude of the sterile neutrino state vanishes if states of massive neutrinos are coherent.
Even in the case when massive neutrinos are in incoherent mixture, the widths  of decays of $W\to \ell_\alpha \nu_i$ and $Z\to \overline{\nu}_i\nu_j$, where $i,j\in (1\dots 4)$, are proportional to the number of active neutrino species (three in the SM), thus avoiding inconsistencies with observations.
This elegant theoretical construction appears when there is a disparity between active and inert numbers of neutrino fields in an extension of the SM.

Currently, there are data consistent with sterile neutrino hypothesis at about 3$\sigma$ with $\Delta m^2=1.3$ eV${}^2$ and $|V_{e4}|^2\approx 0.1$.
On the other hand there is a bulk of accelerator and atmospheric data strongly disfavoring LSND and MiniBooNE anomalies as  due to  sterile neutrino oscillation.
Finally, cosmology also disfavors a fourth neutrino with eV mass and $|V_{\alpha 4}|\simeq 0.1$ -- the only currently available domain of parameters for sterile neutrino hypothesis of reactor $\overline{\nu}_e$ data.

Still, the allowed parameter space is within the sensitivity region of currently running experiments (see Ref.~\cite{Gariazzo:2017fdh} and references therein).

\section{Acknowledgments}
We sincerely thank V.~A.~Naumov and C.~T.~Kullenberg for reading the manuscript and making important comments.

\end{document}